\documentclass[runningheads]{llncs}
\usepackage[T1]{fontenc}
\usepackage{graphicx}
\usepackage{multirow}
\usepackage{threeparttable}
\usepackage{hyperref}
\usepackage{color}

\begin{document}
\title{Agent-Based Modeling of C. Difficile Spread in Hospitals: Assessing Contribution of High-Touch vs. Low-Touch Surfaces and Inoculations' Containment Impact}
\titlerunning{Agent-Based Modeling of C. Difficile Spread}
%
\author{Sina Abdidizaji\inst{1} \and
Ali Khodabandeh Yalabadi\inst{1} \and
Mehdi Yazdani-Jahromi\inst{2} \and
 Ozlem Ozmen Garibay\inst{1} \and 
Ivan Garibay\inst{1}}
\authorrunning{S. Abdidizaji et al.}
%
\institute{Industrial Engineering and Management Systems, University of Central Florida, Orlando FL 32816, USA \and
Computer Science, University of Central Florida, Orlando FL 32816, USA\\
\email{\{sina.abdidizaji, yalabadi, yazdani, ozlem, igaribay\}@ucf.edu}}

\maketitle              
\begin{abstract}
Health issues and pandemics remain paramount concerns in the contemporary era. Clostridioides Difficile Infection (CDI) stands out as a critical healthcare-associated infection with global implications. Effectively understanding the mechanisms of infection dissemination within healthcare units and hospitals is imperative to implement targeted containment measures. In this study, we address the limitations of prior research by Sulyok et al., where they delineated two distinct categories of surfaces as high-touch and low-touch fomites, and subsequently evaluated the viral spread contribution of each surface utilizing mathematical modeling and Ordinary Differential Equations (ODE). Acknowledging the indispensable role of spatial features and heterogeneity in the modeling of hospital and healthcare settings, we employ agent-based modeling to capture new insights. By incorporating spatial considerations and heterogeneous patients, we explore the impact of high-touch and low-touch surfaces on contamination transmission between patients. Furthermore, the study encompasses a comprehensive assessment of various cleaning protocols, with differing intervals and detergent cleaning efficacies, in order to identify the most optimal cleaning strategy and the most important factor amidst the array of alternatives. Our results indicate that, among various factors, the frequency of cleaning intervals is the most critical element for controlling the spread of CDI in a hospital environment.

\end{abstract}
\section{Introduction}
In the contemporary landscape, combating pandemics and health-related challenges necessitates a profound comprehension of contamination transmission and the implementation of appropriate intervention measures to curb infection spread effectively. Clostridioides Difficile Infection (CDI), a critical healthcare-associated infection of global significance \cite{lanzasReviewEpidemiologicalModels2022}, manifests through diverse contaminated means which are capable of contaminating people, such as door handles, knobs, toilets, food, asymptomatic carriers, CDI patients, animals, and environmental sources, each characterized by varying interaction rates. CDI poses a severe threat, leading to diarrhea in patients and, in some cases, fatal outcomes \cite{barkerInterventionsReduceIncidence2018,lanzasReviewEpidemiologicalModels2022}. Consequently, devising novel and efficacious containment strategies remains paramount for healthcare authorities.

Recent work by Sulyok et al. \cite{sulyokMathematicallyModelingEffect2021} has categorized touch surfaces into high-touch and low-touch fomites, developing a mathematical differential equation model to discern the distinct contributions of each surface in the transmission of this specific virus. To assess intervention efficacy in curtailing CDI spread, four types of models have been employed, namely deterministic and stochastic models based on random events, compartmental models tracking population-level changes, and agent-based models tracking individual-level changes. However, real-world interventions can prove prohibitively expensive to implement and measure \cite{lanzasReviewEpidemiologicalModels2022}.

To overcome this challenge, Agent-Based Models (ABMs) present a cost-effective and efficient approach to explore intervention effects within a simulated hospital environment. ABMs offer a powerful means of studying infectious disease propagation, including hospital settings, due to their flexibility and bottom-up approach, allowing the modeling of complex systems and exploring interactions between autonomous and heterogeneous agents to uncover emergent phenomena \cite{bonabeauAgentbasedModelingMethods2002a,epsteinAgentbasedComputationalModels1999}.

In this paper, we replicate Sulyok et al.'s model using agent-based modeling while incorporating heterogeneity and spatial features, leading to a more comprehensive understanding of agent behavior. Additionally, we explore the efficacy of various inoculations on high-touch and low-touch fomites, considering different intervals and detergent cleaning powers. Our research aims to address the following pivotal questions:
\begin{enumerate}
\item How much is the contribution of high-touch and low-touch surfaces in the spread of the virus when heterogeneous agents and spatial features were considered?

\item What are the most effective cleaning strategies given different intervals between each cleaning and different detergent cleaning powers?

\item What is the most important factor in cleaning strategies to contain the spread of the CDI?
\end{enumerate}
The subsequent sections are organized as follows. Section 2 provides a comprehensive review of relevant literature on epidemiological models employed for modeling C. Difficile transmission in hospital settings. In Section 3, we outline the methodology employed in this study. The experimental design, comprising various experiments, is detailed in Section 4. Section 5 presents the results obtained from these experiments, while Section 6 encompasses the concluding remarks drawn from our findings.

\section{Literature Review}
In previous studies, interventions for CDI have been examined, but their simulated environments often lacked the complexity found in real hospital settings. Researchers have attempted to measure the effect of each intervention individually and then bundled them to assess their collective impact on preventing the spread of CDI. For instance, Barker et al. \cite{barkerReducingDifficileChildren2020} investigated interventions such as hand hygiene, contact precautions, daily disinfection, and combinations thereof. In \cite{nelsonEconomicAnalysisStrategies2016}, the researchers also took into account the associated costs of single and bundled interventions to evaluate the cost-effectiveness of disinfection strategies. Among the interventions considered, hand hygiene, environmental disinfection, and isolation and treatment emerged as the most effective and economically viable options. Another study introduced two types of agents in its modeling: animate and inanimate agents \cite{friesenSurveyAgentBasedModeling2014}. Animate agents in this CDI model encompassed patients, visitors, caregivers, nurses, and physicians, while inanimate agents included door handles, door knobs, medical equipment, computers, beds, and other environmental elements \cite{barkerReducingDifficileChildren2020,friesenSurveyAgentBasedModeling2014}.

Numerous studies have investigated various aspects of surface contamination and disinfection in the context of infectious diseases. Abu et al. \cite{abujaradOmniphobicSprayCoating2023} examined the quantity of pathogens present on high-touch surfaces after human contact, shedding light on potential sources of transmission. Conversely, low-touch surfaces, regarded as neglected surfaces in \cite{okaforInvestigatingBioburdenNeglected2022}, were explored to measure the number of microbes on those surfaces, providing insights into areas that might be overlooked in cleaning procedures.

In the context of the SARS-CoV-2 virus, Hardison et al. \cite{hardisonEfficacyChemicalDisinfectants2023} delved into the dynamics of surface materials, disinfectant chemicals, and disinfection protocols, aiming to optimize strategies for virus elimination. Similarly, Nelson et al. \cite{nelsonEfficacyDetergentbasedCleaning2023} investigated cleaning procedures for surfaces, specifically exploring the efficacy of pre-wetting and wiping with hard water in reducing the virus's presence. Their findings revealed that wiping with hard water was effective, whereas pre-wetting surfaces did not significantly impact virus reduction.

The efficiency of cleaning strategies for methicillin-resistant Staphylococcus aureus (MRSA) was assessed by Lei et al. \cite{leiExploringSurfaceCleaning2017} on the hands and noses of two patients, as well as on high-touch and low-touch surfaces. Their study highlighted that the frequency of cleaning played a crucial role, surpassing the significance of cleaning the entire room in mitigating MRSA transmission.

Additionally, Scaria et al. \cite{scariaAssociationVisitorContact2021} employed an Agent-Based Model (ABM) to explore contact between visitors and CDI patients within a hospital setting. Their investigations revealed that interventions focused on hospital workers and surface cleaning were more effective in reducing CDI transmission compared to interventions aimed at reducing visitor-patient contact.

In one of the recent studies by \cite{sulyokMathematicallyModelingEffect2021}, the focus was on fomite surfaces, which can become contaminated and contribute to the transmission of CDI through touch. These fomites were categorized into high-touch and low-touch surfaces. Using ordinary differential equations, the authors estimated that approximately 75-79 percent of contamination spread could be attributed to high-touch fomites, while the remaining 21-25 percent could be attributed to low-touch fomites. Despite the common use of differential equations as a modeling approach, they possess two significant limitations. Firstly, they assume homogeneous agents and overlook individual differences in behavior and susceptibility to infection. Secondly, they fail to consider the spatial dimension of the system, which is particularly crucial in hospital and clinic settings where physical layout and movement can significantly impact disease spread. In light of these limitations, this paper proposes a replication of the findings by \cite{sulyokMathematicallyModelingEffect2021} using an agent-based modeling approach, enhancing the original model with heterogeneity and spatial features to capture the system's behavior more realistically. The use of agent-based modeling has already been explored in studying the spread of the Covid-19 virus in societies and hospitals \cite{daghririQuantifyingEffectsSocial2021,rajabiInvestigatingDynamicsCOVID192021}. Furthermore, our model will investigate the effectiveness of different inoculations and hygienic plans to reduce the spread of CDI. While the effectiveness of inoculations has been explored in previous studies \cite{stephensonComparingInterventionStrategies2020}, our model will uniquely incorporate high-touch and low-touch fomites together by Agent-Based Modeling to measure the contribution of virus spread by each type of surface. 

\section{Methodology}
This paper adopts an extended version of the SIR model, originally introduced by Kermack and McKendrick \cite{kermackContributionMathematicalTheory1997}, to comprehensively simulate the spread of CDI within a hospital environment. To achieve this, we integrate and refine the existing agent-based model proposed by \cite{scariaValidatingAgentbasedSimulation2023} by adding the touch surfaces in \cite{sulyokMathematicallyModelingEffect2021}, effectively incorporating high-touch and low-touch fomites and considering the influence of patients in contaminating these surfaces. Additionally, we investigate the role of various surfaces in the transmission dynamics of the virus, further enhancing the agent-based modeling framework for CDI described in \cite{scariaValidatingAgentbasedSimulation2023}.

In the context of agent-based modeling, it has three fundamental components to define these models: actors, the environment, and interactions between them, all of which play pivotal roles in shaping the simulation and outcome analysis.

\subsection{Actors}
The SIR model, as applied in this study, encompasses three distinct states for each patient: Susceptible, Infected, and Recovered \cite{kermackContributionMathematicalTheory1997}. However, in the context of Clostridioides Difficile Infection, patients may exhibit four distinct states, namely Resistant, Susceptible, Colonized, and Diseased \cite{sulyokMathematicallyModelingEffect2021}. It is noteworthy that in certain literature exploring the spread of CDI, the Resistant state is alternatively referred to as Non-susceptible \cite{scariaValidatingAgentbasedSimulation2023}. Following the definition provided by \cite{sulyokMathematicallyModelingEffect2021}, each state is characterized as follows:
\begin{itemize}
\item Resistant patients (R) refer to individuals who have not undergone any antibiotic treatment, maintain a healthy gut microbiota, and display immunity to C. difficile colonization.
\item Susceptible patients (S) are those who have recently undergone antibiotic treatment, leading to compromised gut microbiota and rendering them vulnerable to C. difficile colonization.
\item Colonized patients (C) are those who harbor C. difficile in their gut without displaying any symptoms of the infection. These patients are asymptomatically infected.
\item Diseased patients (D) are categorized as colonized individuals who exhibit symptomatic manifestations of C. difficile infection. These patients are symptomatically infected.
\end{itemize}
\subsection{Environment}
The environment within this model represents a hypothetical hospital setting, comprising a total of 144 surfaces. These surfaces are further categorized into two distinct groups: high-touch and low-touch surfaces, equally divided in number. Patients, acting as agents in the simulation, are randomly distributed throughout this environment, enabling a comprehensive assessment of infection dynamics and the impact of various cleaning procedures. To ensure fairness and equitable exposure for patients to both high-touch and low-touch fomites, the environment is symmetrically designed.

According to \cite{sulyokMathematicallyModelingEffect2021}, high-touch fomites are found to be interacted with nearly twice as frequently as low-touch surfaces, thus underscoring the significance of their potential role in the transmission of CDI within the hospital environment.
\subsection{Interactions}
\begin{figure}
\includegraphics[width=\textwidth]{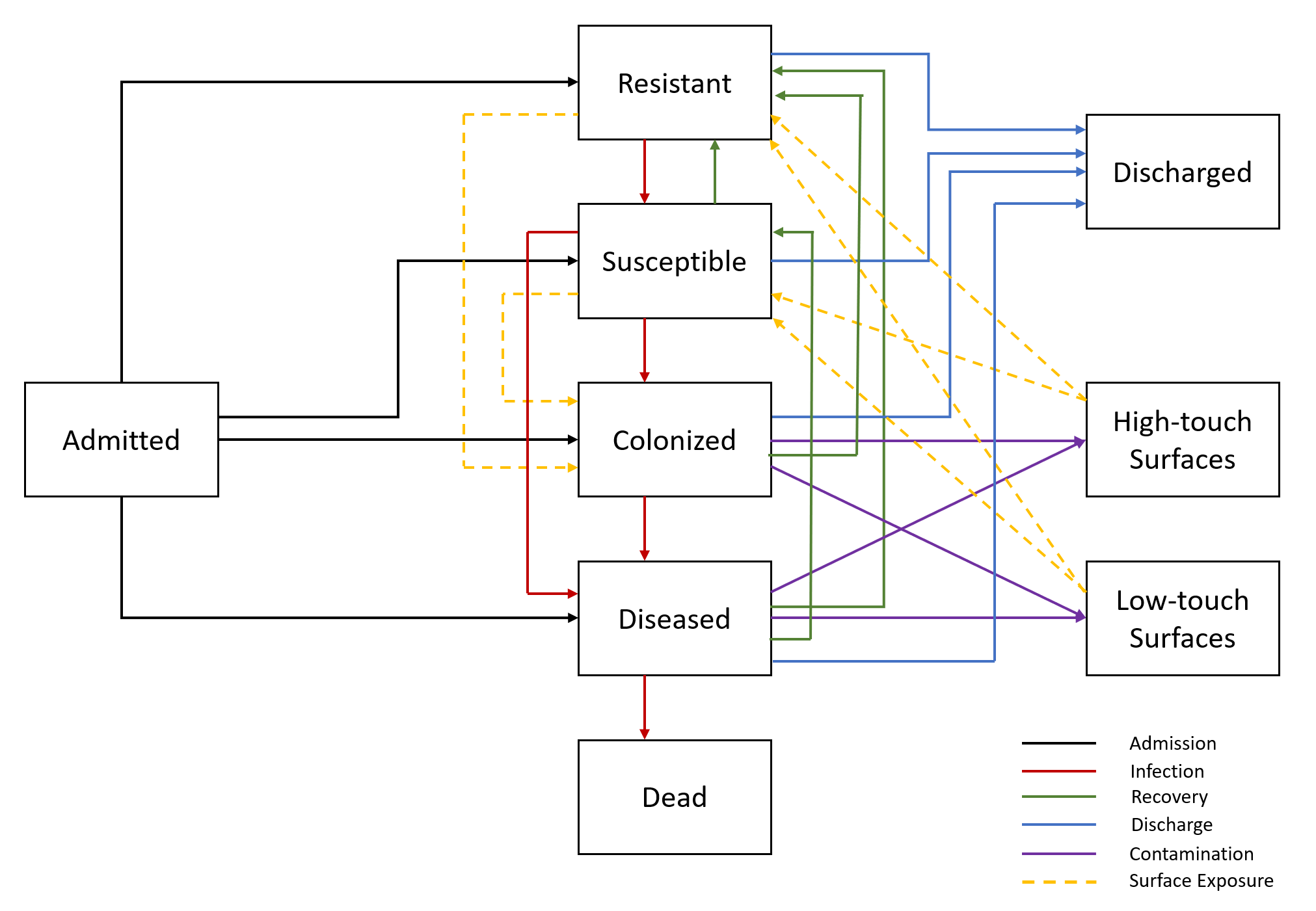}
\caption{The flow of state transmission for a patient in Clostridioides Difficile Infection} \label{fig1}
\end{figure}

In this model, state transitions and surface exposures are governed by interactions between different agents as well as interactions between agents and the environment. The flow of patients and the dynamic transitions between different states are visually depicted in \autoref{fig1}. Upon admission to the hospital, patients may be in a resistant, susceptible, colonized, or diseased state. For those initially in the resistant state, the following scenarios may unfold: after 10 iterations in our model, they may transition to the susceptible state, indicating that they have received antibiotics. If they do not receive antibiotics, they will be discharged from the hospital. During their stay, they may encounter both low-touch and high-touch surfaces, and if they come into contact with highly contaminated surfaces, they could become infected and move to the colonized state.

Patients initially in the susceptible state, after 30 iterations \cite{sulyokMathematicallyModelingEffect2021}, may transition to the resistant, colonized, or diseased state. If they do not develop any issues, they will eventually be discharged from the hospital. Exposure to high-touch and low-touch fomites during their stay may lead to infection and subsequent colonization.

Patients initially in the colonized state, after 15 iterations \cite{crobachUnderstandingClostridiumDifficile2018}, may transition to the diseased or resistant state. Some of them may also be discharged if further treatment is unnecessary. As asymptomatic carriers, colonized patients have a lower rate of contamination shedding compared to those who are diseased. They may contaminate surfaces, both high-touch and low-touch, while touching them.

Patients initially in the diseased state, after 10 iterations \cite{sulyokMathematicallyModelingEffect2021}, may transition to the susceptible state or succumb to death. As previously mentioned, this virus can be fatal, leading to a considerable number of patients experiencing severe infection and mortality \cite{barkerInterventionsReduceIncidence2018}. If no complications arise, they can be discharged from the hospital. During their stay, they have the potential to contaminate the environment and surfaces. The rates of hospital admission, shedding rates, and all other pertinent parameters are provided in \autoref{tab1}.

The initial setup comprises 50 patients randomly placed at different locations within the hospital. Subsequently, at each iteration, 10 new patients are introduced to the hospital. The hospital's capacity is set at 150 patients, and admission is restricted to this limit until previous patients are discharged, creating space for new admissions. The duration of each patient's hospital stay, along with the dynamic transitions between different states, dictates their progression through various states until eventual discharge or, in some unfortunate cases, demise, resulting in their removal from the hospital environment. Patients follow a random walk model for their movements. As the environment in the simulation software consists of squares, each patient can move in 8 directions: four diagonal, two horizontal, and two vertical directions.

\begin{table}[ht]
\caption{Parameters of the model and their descriptions}
\label{tab1}
\centering
\resizebox{\textwidth}{!}{
\begin{tabular}{cccc}
\hline
\textbf{Parameter}& \textbf{Description}& \textbf{Value}& \textbf{Reference} \\ \hline
\textit{$a_R$} & Proportion of patients admitted in a resistant state & 0.75 & \cite{stephensonOptimalControlVaccination2017} \\

\textit{$a_S$} & Proportion of patients admitted in a susceptible state & 0.09 & \cite{stephensonOptimalControlVaccination2017} \\ 

\textit{$a_C$} & Proportion of patients admitted in a colonized state & 0.15 & \cite{stephensonOptimalControlVaccination2017} \\ 

\textit{$a_D$} & Proportion of patients admitted in a diseased state & 0.01 & \cite{stephensonOptimalControlVaccination2017} \\ 

\textit{$k_R$} & Proportion of resistant patients discharged from a hospital & 0.33 & \cite{lanzasEpidemiologicalModelClostridium2011} \\

\textit{$k_S$} & Proportion of susceptible patients discharged from a hospital & 0.15 & \cite{lanzasEpidemiologicalModelClostridium2011} \\ 

\textit{$k_C$} & Proportion of colonized patients discharged from a hospital & 0.15 & \cite{lanzasEpidemiologicalModelClostridium2011} \\ 

\textit{$k_D$} & Proportion of diseased patients discharged from a hospital & 0.068 & \cite{lanzasEpidemiologicalModelClostridium2011} \\

\textit{$P_{R,S}$} & Proportion of resistant patients who become susceptible & 0.02 & \cite{codellaAgentbasedSimulationModel2015} \\

\textit{$P_{S,C}$} & Proportion of susceptible patients who become colonized & 0.001 & \cite{codellaAgentbasedSimulationModel2015} \\

\textit{$P_{S,D}$} & Proportion of susceptible patients who become diseased & 0.001 & \cite{codellaAgentbasedSimulationModel2015} \\

\textit{$P_{C,D}$} & Proportion of colonized patients who become diseased & 0.024 & \cite{stephensonOptimalControlVaccination2017} \\

\textit{$P_{C,R}$} & Proportion of colonized patients who become resistant & 0.012 & \cite{codellaAgentbasedSimulationModel2015} \\

\textit{$P_{D,R}$} & Proportion of diseased patients who become resistant & 0.01 & \cite{codellaAgentbasedSimulationModel2015} \\

\textit{$P_{D,S}$} & Proportion of diseased patients who become susceptible & 0.08 & \cite{mcfarlandUpdateChangingEpidemiology2008} \\

\textit{$P_{Death}$} & Proportion of diseased patients who become dead because of infection & 0.009 & \cite{codellaAgentbasedSimulationModel2015} \\

\textit{$P_{I,C}$} & Probability of become infected by a colonized patient & 0.48 & \cite{codellaAgentbasedSimulationModel2015} \\ 

\textit{$P_{I,D}$} & Probability of become infected by a diseased patient & 0.382 & \cite{codellaAgentbasedSimulationModel2015} \\ 

\textit{$P_{HT}$} & Probability of touching a high-touch surface  & 0.4 & \cite{sulyokMathematicallyModelingEffect2021} \\ 

\textit{$P_{LT}$} & Probability of touching a low-touch surface & 0.2 & \cite{sulyokMathematicallyModelingEffect2021} \\ 

\textit{Colonized Shedding Rate} & The amount of contamination spread by a colonized patient to surfaces & 0.006 & \cite{sulyokMathematicallyModelingEffect2021} \\

\textit{Diseased Shedding Rate} & The amount of contamination spread by a diseased patient to surfaces & 0.013 & \cite{sulyokMathematicallyModelingEffect2021} \\ \hline


\end{tabular}}
\end{table}

\section{Experimental Design}

In order to address three distinct research questions, a series of experiments were devised, and the model was subjected to rigorous testing. Three factorial experiments were conducted, each consisting of 150 replicas, to ensure robustness and statistical validity. The independent variables, their descriptions, and corresponding quantities are outlined in \autoref{tab2}. 
\begin{table}[ht]
\caption{Independent Variables and their values for experimental design}\label{tab2}
\centering
\resizebox{\textwidth}{!}{
\begin{tabular}{cccc} \hline
\textbf{Variable}& \textbf{Description}& \textbf{Values}& \textbf{Experiment}\\ \hline
\textit{direct-infection}   & \begin{tabular}[c]{@{}c@{}}If it is True, there will be direct infection between patients\\ If it is False, there is no direct infection between patients\end{tabular}        & \{True, False\}  & 1                                                                                                  \\ \hline
\textit{disinfect}      & \begin{tabular}[c]{@{}c@{}}If it is True, all of the surfaces will be cleaned\\ If it is False, nothing will be cleaned and there is no cleaning plans\end{tabular}           & \{True, False\} & 2                                                                                                   \\ \hline
\textit{high-touch-disinfection-interval} & \begin{tabular}[c]{@{}c@{}}This illustrates the cleaning frequency of a high-touch surface\end{tabular}                                                                    & \{10, 20, 30\}  & 2, 3                                                                                                     \\ \hline
\textit{high-touch-disinfection-rate}      & \begin{tabular}[c]{@{}c@{}}This shows the potency of a detergent for cleaning high-touch surfaces \end{tabular}
                            & \{0.5, 0.7, 0.9, 1\} & 2, 3  
                                                     \\ \hline
\textit{low-touch-disinfection-interval} & \begin{tabular}[c]{@{}c@{}}This illustrates the cleaning frequency of a low-touch surface\end{tabular}                                                                    & \{10, 20, 30\}  & 2, 3                                                                                                       \\ \hline
\textit{low-touch-disinfection-rate}      & \begin{tabular}[c]{@{}c@{}}This shows the potency of a detergent for cleaning low-touch surfaces \end{tabular}
                            & \{0.5, 0.7, 0.9, 1\} & 2, 3  
                                                    \\ \hline
\textit{random-disinfect}      & \begin{tabular}[c]{@{}c@{}}If it is True, some of the surfaces will be cleaned randomly\\ If it is False, nothing will be cleaned and there is no cleaning plans\end{tabular} & \{True, False\} & 3                                                                                                     \\ \hline
\textit{number-of-random cleaning}      & \begin{tabular}[c]{@{}c@{}}This denotes the number of surfaces from each type to be cleaned randomly\end{tabular}                                                            & \{24, 36, 48\} & 3                                                                                                           \\ \hline
\end{tabular}}
\end{table}

\subsection{Experiment 1}
The primary objective of this experiment was to quantify the contributions of high-touch and low-touch fomites to the transmission of the C. difficile virus within a hospital setting. Initially, the assumption was made that patients could only contract the infection through surface contact, without considering the possibility of infection via exposure to colonized or diseased patients. Subsequently, the model was reevaluated to explore the impact of direct infection from other patients, revealing potential differences in surface contributions to contamination under this scenario. For these experiments, no cleaning plans were factored into the simulation, and any contaminated surface remained so throughout the duration of the patient's stay in the hospital.
\subsection{Experiment 2}
Following the determination of each surface's contribution to contamination, the subsequent experiment focuses on introducing cleaning plans to evaluate their effectiveness in containing the spread of contamination. For each cleaning plan, two distinct variables are considered: the interval between each cleaning session and the power of the detergent employed during the cleaning process. The interval between cleanings is set using steps in the simulation software. For instance, if we set it to 10, it means that every 10 simulation cycles, all the surfaces get cleaned. The detergent's efficacy is probabilistically defined, with a value of 0.5 indicating that only 50 percent of the contamination on a surface will be removed upon cleaning. Higher values signify more potent cleaning detergents. Additionally, all surfaces, whether high-touch or low-touch, are subjected to cleaning and disinfection simultaneously.

Notably, in this particular experiment, direct infection between patients is not considered; the infection transmission is solely dependent on patients' contact with high-touch and low-touch fomites.
\subsection{Experiment 3}
This experiment closely resembles the previous one, with a minor distinction in the cleaning protocol. In this scenario, not all surfaces undergo simultaneous cleaning; instead, the cleaning process occurs randomly. This means that during each cleaning cycle, only some surfaces are randomly selected and cleaned. As a result, some surfaces might be cleaned in two consecutive cleaning processes, while others may not be cleaned at all, even after five cleaning cycles, because the selection is entirely random. This assumption reflects real-world conditions where limited cleaning personnel are unable to attend to all surfaces simultaneously, resulting in some surfaces being cleaned regularly while others may not receive immediate attention.

Importantly, similar to the previous experiment, direct infection between infected patients is not considered in this study. Rather, the sole mode of infection transmission remains through patients' contact with contaminated surfaces.

The model utilized in this study was implemented using NetLogo, a widely recognized agent-based modeling platform \cite{Tisue2004NetLogoAS,wilenskyIntroductionAgentBasedModeling2015} while the statistical analysis was performed using Python. For reference, Appendix provides a detailed view of the initial model interface within the NetLogo software, encompassing all relevant parameters used in the simulations.

\section{Results}
\subsection{Direct Contact vs. Surface Contact}
The experiments comprised a total of 86,700 simulations conducted under diverse conditions. In Experiment 1, the absence of any cleaning plans resulted in gradual contamination of all surfaces to varying degrees. We examined the dynamic spread of contamination in scenarios involving both direct contact infection and its absence. \autoref{tab3} presents the average number of individuals infected through surface contact and direct contact with infected patients.  Given the stochastic nature of the model, the average number of infected patients from both high-touch and low-touch surfaces was reported in tables and figures in this section.
\begin{table}[ht]
\caption{The average contribution of surfaces and patients in infecting resistant and susceptible patients}\label{tab3}
\centering
\resizebox{\textwidth}{!}{
\begin{tabular}{|c|c|c|c|c|c|c|} \hline
Experiment & Surface Infection & Direct Contact Infection & High-touch & Low-touch & Colonized Patient & Diseased Patient \\ \hline
1          & True              & True                     & 13\%         & 7.3\%       & 69.7\%              & 10\%               \\
1          & True              & False                    & 64.2\%       & 35.7\%      & -                 & -                \\
2 \& 3     & True              & False                    & 75\%         & 25\%        & -                 & -                \\ \hline
\end{tabular}}
\end{table}

When considering the possibility of direct contact infection from patients, surfaces demonstrated a relatively minor contribution to the infection of both resistant and susceptible patients, accounting for only around 20 percent of the infections as a result of exposure to contaminated fomites. In the absence of direct contact infection, high-touch fomites were responsible for approximately 64 percent of infections, while low-touch fomites contributed to approximately 36 percent. Even when direct contact infection was present, the proportional contributions of each type of surface remained quite similar.

Upon implementing cleaning procedures in our model, the contribution of high-touch and low-touch fomites to contamination changed to an average of approximately 75 percent and 25 percent, respectively. This aligns closely with the results reported by \cite{sulyokMathematicallyModelingEffect2021}, where they found that low-touch fomites contributed around 21 to 25 percent and high-touch fomites contributed around 75 to 79 percent to contamination. The congruence of our results with theirs, considering the inclusion of spatial features and heterogeneity, validates their findings.

These findings underscore the importance of prioritizing the disinfection of high-touch surfaces, as they significantly contribute to contamination spread. Conversely, cleaning low-touch surfaces effectively reduces their contribution by an average of 9 percent. Thus, emphasizing disinfection measures for high-touch surfaces and implementing cleaning strategies for low-touch surfaces can be instrumental in controlling C. difficile contamination.
\subsection{The exploration of factors in surface disinfection}
The cleaning of surfaces was investigated with regard to three distinct factors: the potency of the disinfectant, the method of surface cleaning (i.e., cleaning all surfaces versus cleaning them randomly), and the frequency of cleaning as determined by the interval between each disinfection. We thoroughly examined each of these factors to discern potential differences in their impact.

\subsubsection{Disinfectant Potency}
\begin{figure}
\includegraphics[width=\textwidth]{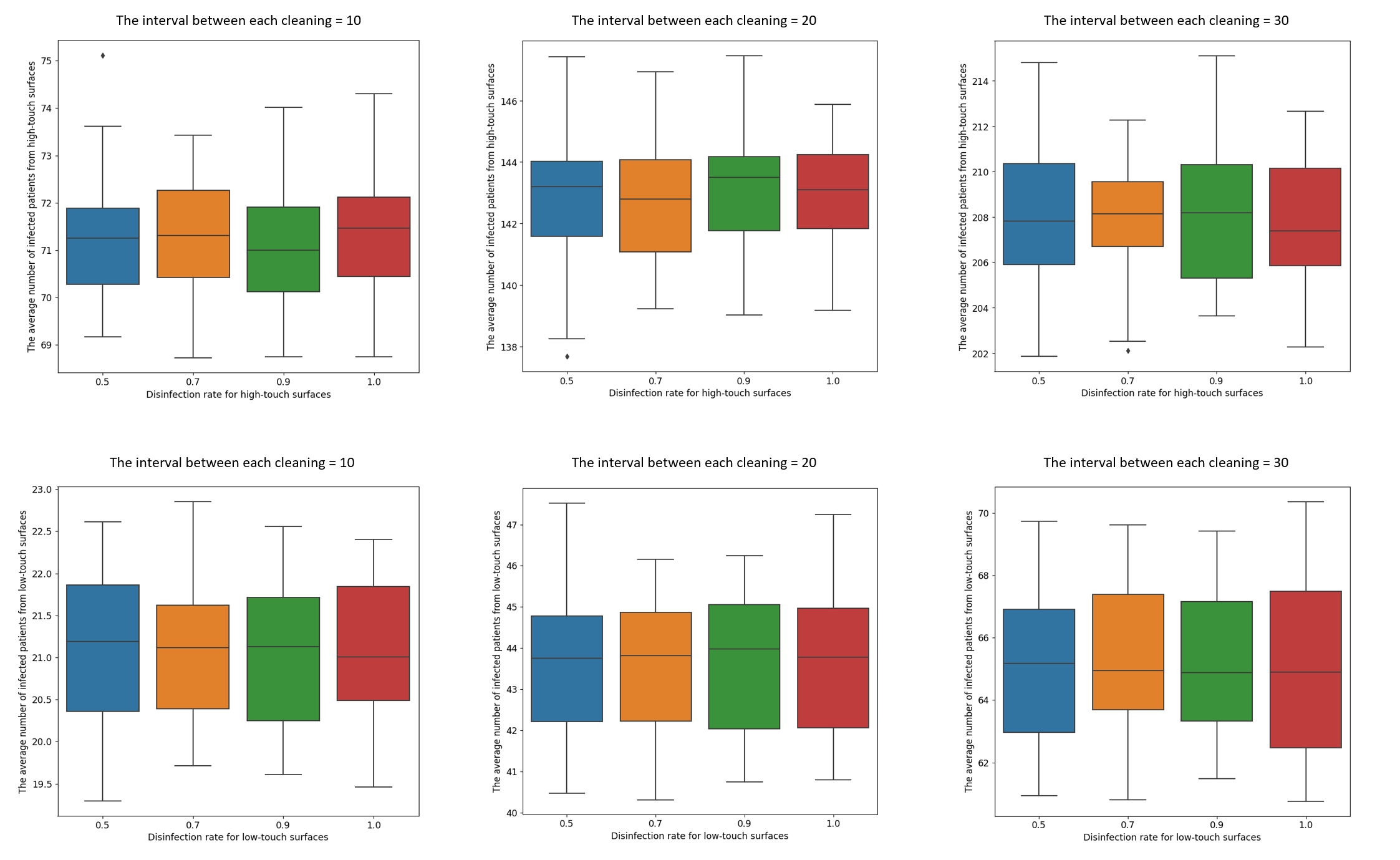}
\caption{The effect of different disinfectants with diverse potency on containing the virus contamination. Different colors represent different potency of disinfectants. Blue for 50 percent potency, Orange for 70 percent potency, Green for 90 percent potency, and Red for 100 percent potency} \label{fig2}
\end{figure}

As indicated in \autoref{tab2}, Experiments 2 and 3 incorporated four different levels of disinfectant potency. Each experiment comprised 150 replicas, with the average contributions of high-touch and low-touch surfaces to the spread of the virus being calculated. The results were represented in boxplots, graphically displayed in \autoref{fig2}.

Upon analyzing the data, it became evident that, for the same cleaning frequency, different levels of cleaning potency did not significantly affect the amount of contamination contribution. Consequently, the potency of the disinfectant may not exert as substantial an influence on the containment of contamination as initially anticipated.
\subsubsection{Total Cleaning vs. Random Cleaning}
\begin{figure}
\includegraphics[width=\textwidth]{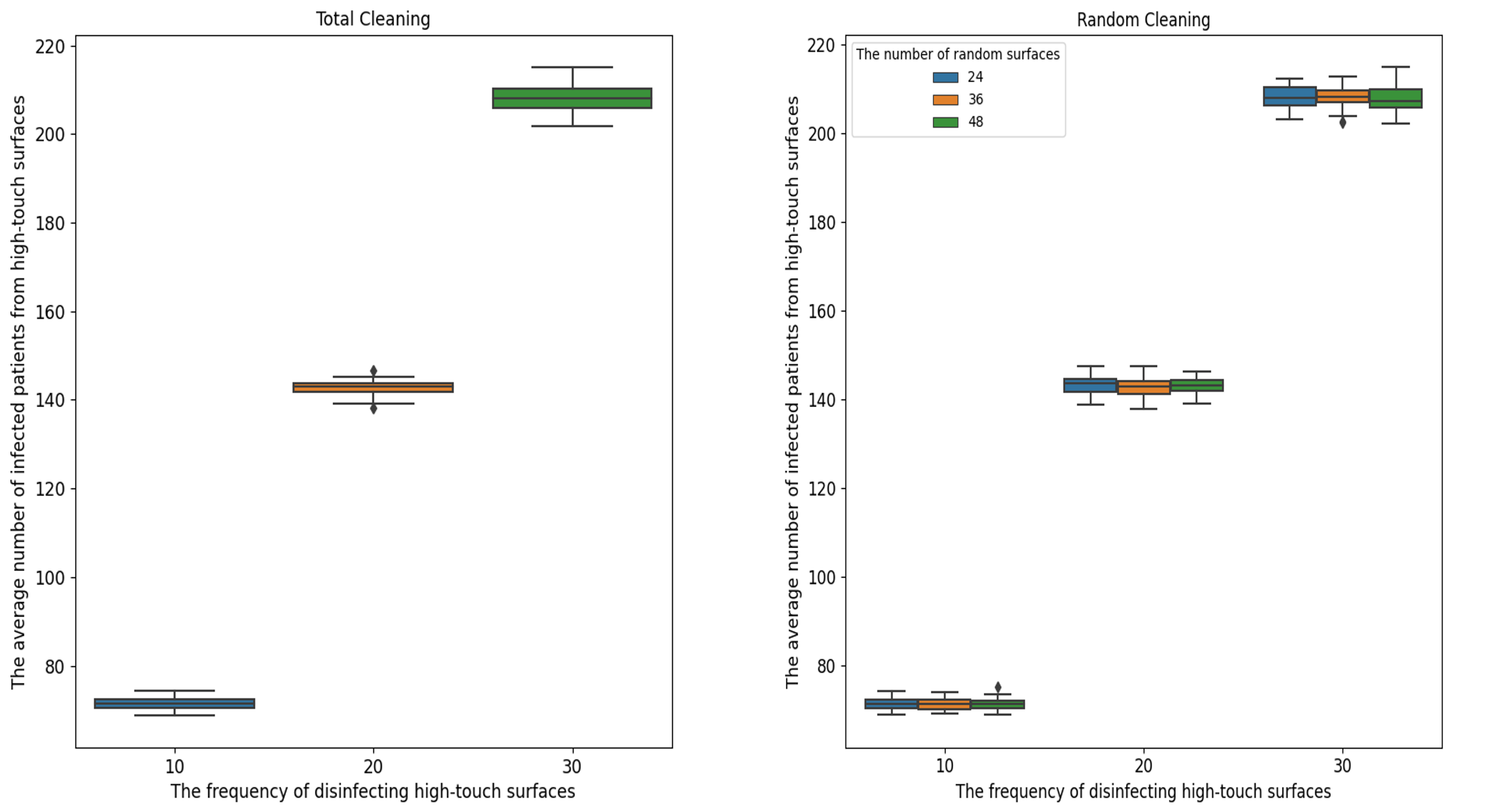}
\caption{Comparing the effect of total cleaning vs random cleaning on high-touch surfaces} \label{fig3}
\end{figure}

\begin{figure}
\includegraphics[width=\textwidth]{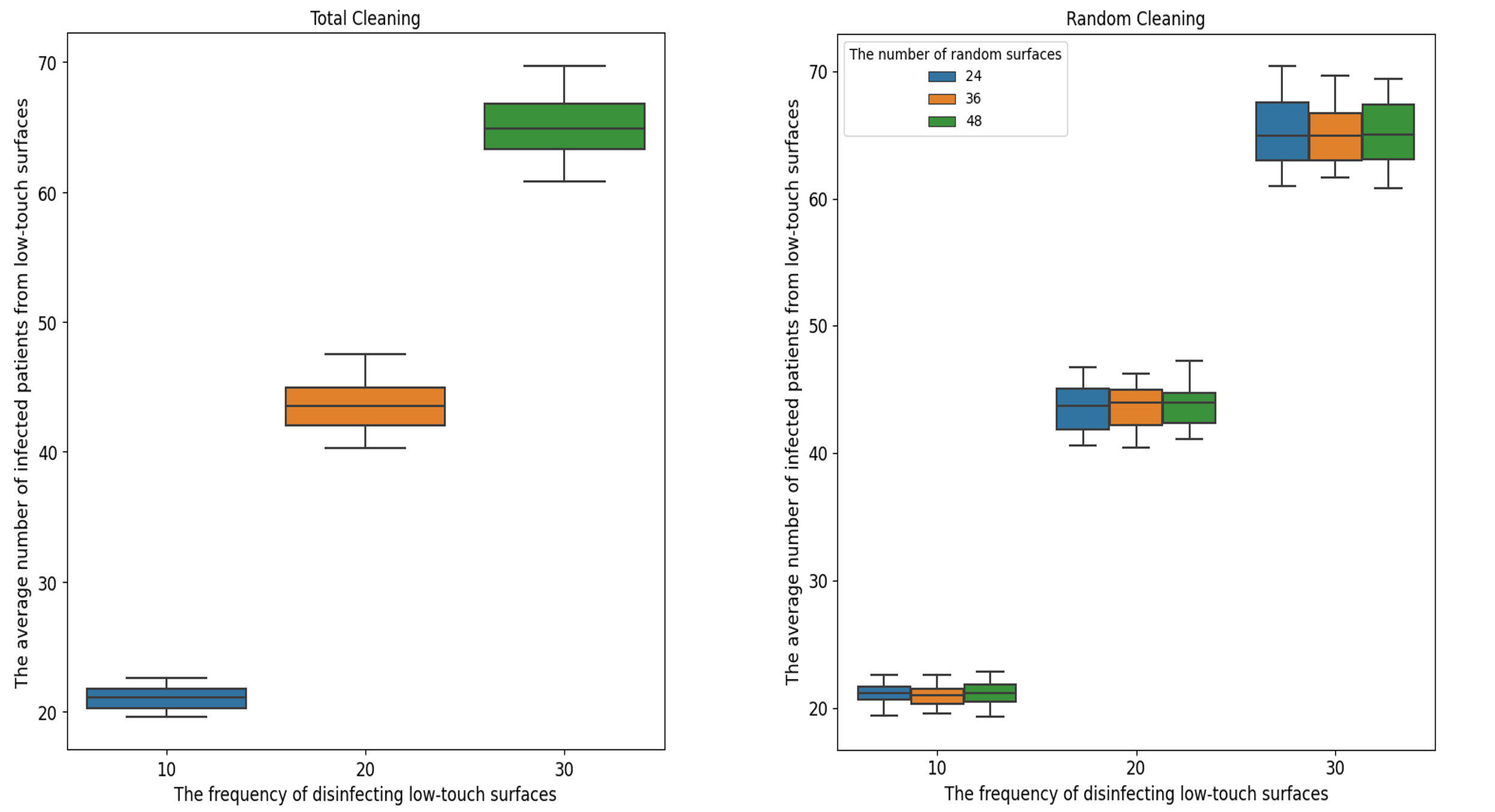}
\caption{Comparing the effect of total cleaning vs random cleaning on low-touch surfaces} \label{fig4}
\end{figure}

Due to limitations in available human workers, not all surfaces can be cleaned with the same frequency, necessitating variations in cleaning schedules. We investigate the effectiveness of total cleaning compared to random cleaning to assess the extent of its impact on contamination contribution. The contamination contributions attributed to high-touch and low-touch fomites in four different scenarios are visually depicted in \autoref{fig3} and \autoref{fig4}, respectively.

As evident from the results, the choice between total cleaning and random cleaning, at the same cleaning interval, did not significantly affect the containment of virus spread. For instance, in the case of high-touch fomites, when the cleaning frequency occurred every 20 iterations, the amount of virus spread ranged from 135 to 150. Consequently, this factor did not demonstrate a substantial role in the efficacy of surface cleaning.

\subsubsection{Cleaning Frequency}
The final factor examined in this study is the cleaning frequency. Through careful examination of \autoref{fig3} and \autoref{fig4}, it becomes evident that altering the cleaning intervals and frequency leads to a dramatic change in the contamination contribution by both high-touch and low-touch surfaces. Notably, the contamination contribution increases two times and three times when the cleaning intervals are changed from 10 to 20 and 10 to 30, respectively. Furthermore, \autoref{fig2} also demonstrates that different cleaning intervals result in significant variations in contamination contribution.

Thus, the most crucial aspect of surface cleaning appears to be the cleaning frequency. Irrespective of the potency of the detergent and disinfectant employed or the availability of human workers, it is the frequency of cleaning that plays a pivotal role in effectively containing the C. difficile virus. Both high-touch and low-touch surfaces necessitate frequent cleaning, and neither should be overlooked in favor of the other. The findings from the study by \cite{leiExploringSurfaceCleaning2017} align closely with our own results. In their research focusing on MRSA, Lei et al. employed Ordinary Differential Equations (ODE) to measure the efficiency of cleaning procedures and concluded that the frequency of cleaning carries more significance than cleaning the entire room for patients. This parallel outcome fortifies our observations regarding CDI and underscores the critical role of regular and consistent cleaning practices in reducing infection transmission in healthcare settings.
\subsection{The Best Possible Strategy}
The findings of this study revealed that higher cleaning frequency resulted in reduced infection spread among patients who had contact with surfaces. In order to identify the most effective cleaning strategy, various disinfection potencies were compared when the cleaning interval is 10 iterations.

\begin{table}[]
\caption{The average contribution of surfaces to the spread of CDI when cleaning plans were introduced}\label{tab4}
\centering
\resizebox{\textwidth}{!}{
\begin{tabular}{|c|c|c|c|c|c|c|} \hline
Type of Cleaning & Disinfectant Rate & Infected by HT & Infected by LT & Total Infection & Percentage of infection from HT & Percentage of infection from LT \\ \hline
\multirow{4}{*}{Total}               & 0.5                                   & 71.4                               & 20.9                               & 92.3                                & 77.35\%                                             & 22.64\%                                             \\
                                     & 0.7                                   & 70.4                               & 20.3                               & 90.7                                & 77.65\%                                             & 22.34\%                                             \\
                                     & 0.9                                   & 68.8                               & 20                                 & 88.8                                & 77.45\%                                             & 22.54\%                                             \\
                                     & 1                                     & 70.6                               & 19.9                               & 90.5                                & 77.97\%                                             & 22.02\%                                             \\ \hline
\multirow{4}{*}{Random 24}           & 0.5                                   & 69.8                               & 20.2                               & 90                                  & 77.56\%                                             & 22.43\%                                             \\
                                     & 0.7                                   & 70.4                               & 19.8                               & 90.2                                & 78.04\%                                             & 21.95\%                                             \\
                                     & 0.9                                   & 69.3                               & 19.6                               & 88.9                                & 77.92\%                                             & 22.07\%                                             \\
                                     & 1                                     & 72.4                               & 20.2                               & 92.6                                & 78.21\%                                             & 21.78\%                                             \\ \hline
\multirow{4}{*}{Random 36}           & 0.5                                   & 62.8                               & 20.4                               & 90.2                                & 77.41\%                                             & 22.58\%                                             \\
                                     & 0.7                                   & 70                                 & 20.4                               & 90.4                                & 77.45\%                                             & 22.55\%                                             \\
                                     & 0.9                                   & 69.9                               & 20.1                               & 90                                  & 77.62\%                                             & 22.38\%                                             \\
                                     & 1                                     & 70                                 & 20.2                               & 90.2                                & 77.59\%                                             & 22.41\%                                             \\ \hline
\multirow{4}{*}{Random 48}           & 0.5                                   & 70.2                               & 20.3                               & 90.5                                & 77.57\%                                             & 22.43\%                                             \\
                                     & 0.7                                   & 70.8                               & 20.3                               & 91.1                                & 77.69\%                                             & 22.31\%                                             \\
                                     & 0.9                                   & 71.2                               & 20.5                               & 91.7                                & 77.63\%                                             & 22.37\%                                             \\
                                     & 1                                     & 68.7                               & 20.5                               & 89.2                                & 77.06\%                                             & 22.94\%                                             \\ \hline
-                                    & -                                     &     659.6                               &        367.7                            &         1027.3                            &                              64.21\%                       & 35.79\%  \\ \hline                                                 
\end{tabular}}
\begin{tablenotes}
\item \begin{flushleft}
\scriptsize HT = High-touch, LT = Low-touch, Random 24 = 24 surfaces of each type were cleaned randomly, Random 36 = 36 surfaces of each type were cleaned randomly, Random 48 = 48 surfaces of each type were cleaned randomly
\end{flushleft}
\end{tablenotes}
\end{table}

Upon examining \autoref{tab4}, two key insights emerged. Firstly, the introduction of inoculation through disinfecting touch surfaces in the hospital environment led to a significant reduction of over 90 percent in the number of infected patients. Secondly, although different disinfectant potencies did not yield a substantial decrease in the number of infected patients, the most favorable strategies involved the use of potent detergents capable of thoroughly cleaning surfaces, achieving either complete or at least 90 percent elimination of contaminants.

As complete eradication of all germs and viruses might not be practically achievable in real-world settings, attaining a 90 percent reduction in surface contamination can be considered a highly effective and feasible approach. Thus, implementing frequent cleaning regimes and, to some extent, utilizing powerful disinfectants can significantly contribute to limiting the transmission of infections in healthcare environments.

\section{Conclusion}
This paper investigates the significant role of touch surfaces in the transmission of the C. difficile virus. Additionally, we introduce and assess two different classes of cleaning plans, given the impact of cleaning frequency and disinfectant potency. Our findings align with those of Sulyok et al. \cite{sulyokMathematicallyModelingEffect2021}, as we observe that high-touch surfaces on average contribute approximately 75 percent, while low-touch surfaces contribute around 25 percent to the spread of the virus.

Furthermore, our study delves into the effects of three key factors within the cleaning plans. Remarkably, we find that the interval between each cleaning session exerts a dramatic influence on the effectiveness of containment. In essence, regular cleaning proves to be the most crucial factor in controlling the spread of the C. difficile virus. On the other hand, disinfectant potency and cleaning all surfaces simultaneously do not significantly impact the containment of the virus.

Our research emphasizes the critical importance of frequent and consistent cleaning practices in mitigating the spread of CDI. By focusing on regular cleaning intervals, healthcare facilities can enhance their infection control measures and reduce the risk of transmission via touch surfaces effectively.

\subsubsection{Acknowledgements} We express our sincere gratitude to Chathura Jayalath for his invaluable insights and contributions, which significantly enriched the quality of this paper.

%
%

\bibliographystyle{splncs04}
\bibliography{main}

\subsubsection{Appendix} The figure below illustrates the initial configuration of the model, as implemented in the NetLogo software, featuring the corresponding initial parameter values. In this visualization, the cyan squares symbolize the high-touch surfaces, while the pink squares denote the low-touch surfaces. Additionally, the yellow blocks are designated as walls, creating impassable barriers through which patients cannot traverse. The code and ODD protocol document can be found in this \href{https://www.comses.net/codebase-release/fc563fb8-217a-427f-bf5d-57224a7cb35d/}{CoMSES link}.
\begin{figure}
\includegraphics[width=\textwidth]{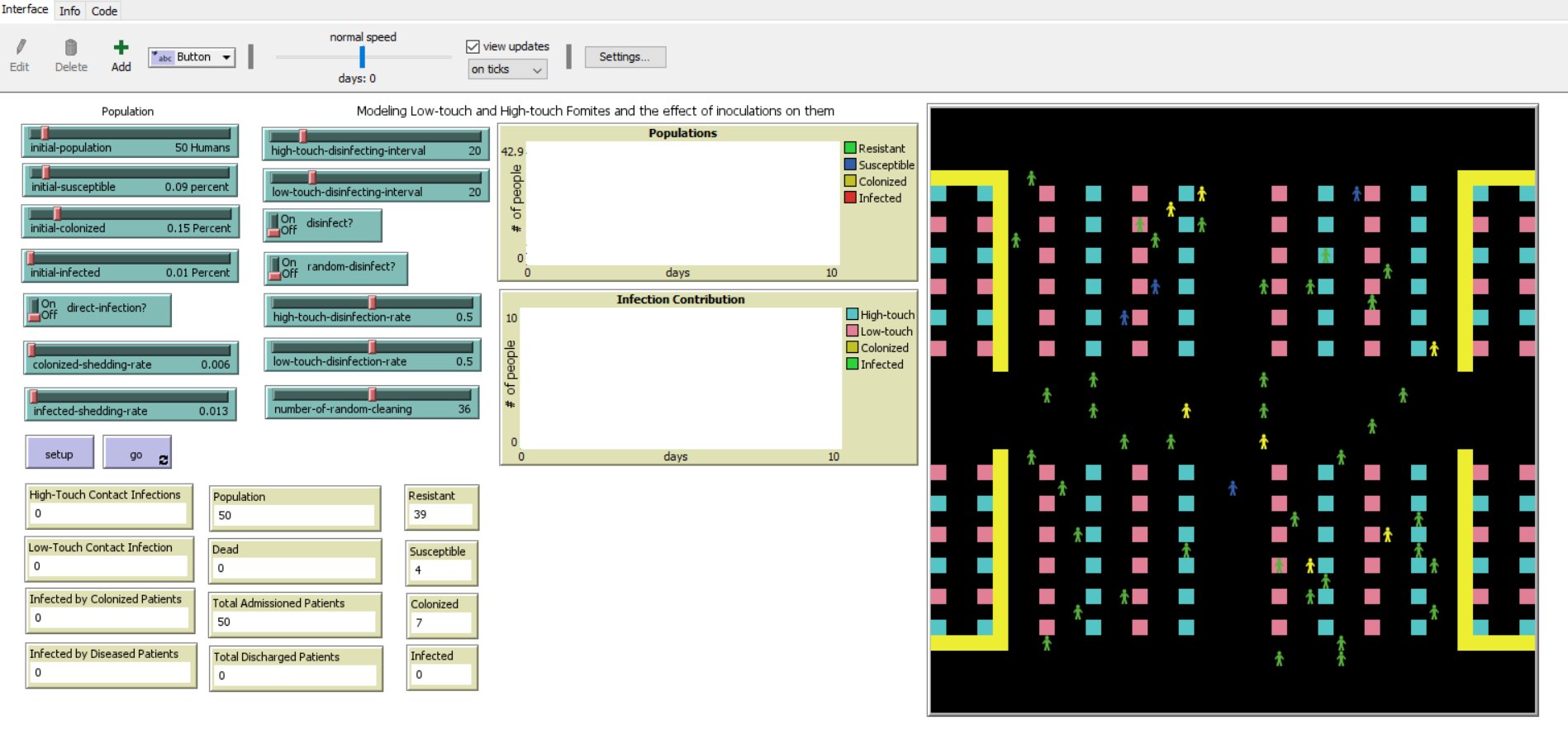}
\end{figure}



\end{document}